\documentclass[aps,twocolumn,a4paper,10pt]{revtex4-1}
\usepackage{mathrsfs}
\usepackage{graphicx}
\usepackage{latexsym}
\usepackage{amsmath}
\usepackage{amssymb}
\usepackage{textcomp}
\usepackage{amsbsy}
\usepackage{color}
\usepackage{epstopdf}

\begin{document}
\title{\large \bf Inflationary scenario from higher curvature warped spacetime}
 \author{Narayan Banerjee}
\email{narayan@iiserkol.ac.in}
\affiliation{Department of Physical Sciences, Indian Institute of Science Education and Research Kolkata,
Mohanpur Campus, Nadia, West Bengal 741246, India}
\author{Tanmoy Paul }
\email{pul.tnmy9@gmail.com}
\affiliation{Department of Theoretical Physics,\\
Indian Association for the Cultivation of Science,\\
2A $\&$ 2B Raja S.C. Mullick Road,\\
Kolkata - 700 032, India.\\}

\begin{abstract}
We consider a five dimensional AdS spacetime, in presence of higher curvature term like 
$F(R) = R + \alpha R^2$ in the bulk, in the context of Randall-Sundrum two-brane model. 
Our universe is identified with the TeV scale brane and emerges 
as a four dimensional effective theory. From the perspective of this effective theory, we examine the 
possibility of ``inflationary scenario'' by considering the on-brane metric ansatz as an FRW one. 
Our results reveal that the higher curvature term in the five dimensional bulk spacetime 
generates a potential term for the radion field. Due to the presence of radion potential, 
the very early universe undergoes a stage of accelerated expansion and moreover the accelerating 
period of the universe terminates in a finite time. We also find the spectral index of curvature perturbation ($n_s$) and 
tensor to scalar ratio ($r$) in the present context, which match with the observational results based on the 
observations of $Planck$ $2015$ \cite{Planck_result}.

\end{abstract}
\maketitle

\section{Introduction}
Over the last two decades, extra spatial dimensions \cite{arkani,horava,RS,kaloper,cohen,burgess,chodos}  has been increasingly playing 
a central role in physics beyond the standard model of particle \cite{rattazzi} and cosmology \cite{marteens}. 
Apart from phenomenological approach, higher dimensional scenarios come naturally in string theory. 
Depending on geometry, the extra dimensions are compactified under various compactification schemes. 
Our usual four dimensional universe is considered to be a 3-brane ($3+1$ dimensional brane) embedded within the higher dimensional 
spacetime and emerges as a four dimensional effective theory.\\
Among the various extra dimensional models proposed over the last several years, Randall-Sundrum (RS) 
warped extra dimensional model \cite{RS} earned a special attention since it resolves the gauge hierarchy 
problem without introducing any intermediate scale (between Planck and TeV scale) in the theory. 
RS model is a five dimensional AdS spacetime with $S^1/Z_2$ orbifolding along the extra dimension 
while the orbifold fixed points are identified with two 3-branes. The separation between the branes 
is assumed to be of the order of Planck length so that the hierarchy problem can be solved. However, 
due to the intervening gravity, aforementioned brane configuration can not be a stable one. 
So, like other higher dimensional braneworld scenario, one of the crucial aspects of RS 
model is to stabilize the interbrane separation (known as modulus or radion). For this 
purpose, one needs to generate a suitable radion potential with a stable minimum. Goldberger 
and Wise (GW) proposed a useful mechanism \cite{GW} to construct such a radion potential 
by imposing a massive scalar field in the bulk with appropriate boundary conditions. Subsequently 
the phenomenology of radion field has also been studied extensively in \cite{GW_radion,csaki,julien,wolfe}.\\ 
Some variants of RS model and its modulus stabilization have been discussed in \cite{csaki,julien,wolfe,ssg1,tp1,tp3,tp2}.\\
The Standard Big Bang model gives a predictive description of our universe from nucleosynthesis to present. 
But back in very early stage of evolution, the big bang model is plagued with some problems such as Horizon and Flatness problems. 
For a comprehensive review, we refer to \cite{perkins,watson}. 
In order to resolve these problems, the idea of inflation was introduced by Guth \cite{guth} in which the universe 
had to go through a stage of accelerated expansion after the big bang. It had also been demonstrated that
a massive scalar field with a suitable potential plays a crucial role in producing an accelerated expansion of the universe. 
This resulted in a huge amount of work on inflation based on scalar fields \cite{perkins, watson,linde,kinney,langlois,habib,riotto,
nb1,barrow1,barrow2,mimiso}. \\
It is interesting to note that in the extra dimensional models, the modulus field can fulfill the requirement of the scalar field 
required for inflation. Thus the cosmology of higher dimensional models \cite{cos1,cos2,cos3,cos4,cos5,cos6,cos7,cos8} can be very different 
from usual cosmology of four dimensions where the inflaton field is normally invoked by handz. In our current work, we take 
advantage of the modulus field of extra dimensions and address the early time cosmology of our universe in the backdrop of RS 
two-brane model.\\
It is well known that Einstein-Hilbert action can be generalized
by adding higher order curvature terms 
which naturally arise from the diffeomorphism property of the action. 
Such terms also have their origin in String Theory due to 
quantum corrections. 
$F(R)$ \cite{faraoni,felice,paliathanasis,nojiri1}, Gauss-Bonnet (GB) \cite{nojiri2,nojiri3,cognola}
or more generally Lanczos-Lovelock gravity \cite{lanczos,lovelock} 
are some of the candidates in higher curvature gravitational 
theory.\\
Higher curvature terms become extremely relevant at the regime 
of large curvature. Thus for RS bulk geometry, where the curvature is of the  
order of Planck scale, the higher curvature terms should play 
a crucial role. Motivated by this idea, we consider a generalized version of 
RS model by replacing Einstein-Hilbert bulk gravity Lagrangian, given by the Ricci scalar $R$ by 
$F(R)$ where $F(R)$ is an analytic function of $R$\cite{marino,bahamonde,catena}. 
Recently it has been shown in \cite{tp1}, that for RS braneworld modified 
by $F(R)$ gravity, a potential term for the radion field is generated (in the 
four dimensional effective theory) even without introducing an external scalar field 
in the bulk and moreover the radion potential has a stable minimum for a certain 
range of parametric space. However, from cosmological aspect, the important questions 
that remain in the said higher curvature RS model \cite{tp1}, are:
\begin{enumerate}
 \item Can the usual four dimensional universe undergo an accelerating expansion 
 at early epoch, due to presence of the radion potential generated by higher curvature term?
 
 \item If such an inflationary scenario is allowed, then what are the dependence of duration of inflation as 
 well as number of e-foldings on higher curvature parameter? Moreover what are the values of $n_s$ and $r$ in the present context?
\end{enumerate}
We aim to address these questions in this work and motivated by the Starobinsky model \cite{starobinsky}, the form of 
$F(R)$ in the five dimensional bulk, is taken here, as $F(R)=R+\alpha R^2$ where $\alpha$ is a constant.\\
The paper is organized as follows: Following two sections are devoted to brief reviews of RS 
scenario and its extension to $F(R)$ model. Section IV is reserved for determining the solutions of 
effective Friedmann equations on the brane. In section V, VI, and VII, we address the consequences 
of the solutions that are obtained in section IV. Finally the paper ends with some concluding  
remarks in section VIII.

\section{Brief description of RS scenario}

RS scenario is defined on a five dimensional AdS spacetime involving one warped and compact 
extra spacelike dimension. Two 3-branes known as TeV/visible and Planck/hidden brane 
are embedded in a five dimensional spacetime. If $\phi$ is the extra dimensional angular 
coordinate, then the branes are 
located at two fixed points $\phi=(0,\pi)$ while the latter one is identified 
with our known four dimensional universe. The opposite brane tensions along with the finely 
tuned five dimensional cosmological constant serve as energy-momentum tensor of RS 
scenario. The resulting spacetime metric \cite{RS} is non-factorizable and expressed as,
\begin{equation}
ds^2 = e^{- 2 kr_c|\phi|} \eta_{\mu\nu} dx^{\mu} dx^{\nu} -r_c^2d\phi^2 \label{eq1}
\end{equation}
Here, $r_c$ is the compactification radius of the extra dimension. 
Due to $S^1/Z_2$ compactification along the extra dimension, $\phi$ ranges from 
$-\pi$ to $+\pi$.
The quantity $k=\sqrt{\frac{-\Lambda}{12M^3}}$, is of the order of 5-dimensional Planck
scale $M$. Thus $k$ relates the 5D Planck scale $M$ to the 5D cosmological constant
$\Lambda$.\\
In order to solve the hierarchy problem, it is assumed in RS scenario that the branes 
are separated by such a distance that 
$k\pi r_c \approx 36$. Then the exponential factor present
in the metric, which is often called warp factor, produces 
a large suppression so that a mass scale of the order of Planck scale is reduced to TeV scale on the 
visible brane. 
A scalar mass e.g. mass of Higgs boson is given as, 
\begin{equation}
 m_H=m_{0}e^{-k\pi r_c}\label{physmass}
\end{equation}
where $m_H$ and $m_0$ are physical and bare Higgs masses respectively.

\section{RS like spacetime in F(R) model: Four dimensional effective action}
In the present paper, we consider a five dimensional AdS spacetime with two 3-brane scenario in F(R) model. The form of 
$F(R)$ is taken as $F(R) = R + \alpha R^2$ where $\alpha$ is a constant with square of the 
inverse mass dimension. Considering $\phi$ as the extra dimensional 
angular coordinate, two branes are located at $\phi = 0$ (hidden brane) and at $\phi = \pi$ (visible brane) respectively 
while the latter one is identified with the visible universe. Moreover the spacetime is $S^1/Z_2$ 
orbifolded along the coordinate $\phi$. 
The action for this model is :
\begin{eqnarray}
 S&=&\int d^4x d\phi \sqrt{G} [\frac{1}{2\kappa^2}(R + \alpha R^2) + \Lambda\nonumber\\ 
 &+&V_h\delta(\phi) + V_v\delta(\phi-\pi)]
 \label{actionF(R)}
\end{eqnarray}
where $G$ is determinant of the five dimensional metric ($G_{MN}$), $\Lambda (< 0)$ is the bulk cosmological constant, 
$\frac{1}{2\kappa^2}=M^3$ and $V_h$, $V_v$ are the brane tensions on hidden, visible brane 
respectively.\\
It is well known that a $F(R)$ gravity model can be recast into Einstein gravity with 
a scalar field by means of a conformal transformation on the metric \cite{nojiri1,tp1}. Thus the solutions of five 
dimensional Einstein equations for the action presented in eqn. (\ref{actionF(R)}), can 
be extracted from the solutions of the corresponding conformally related scalar-tensor (ST) theory 
and it is discussed in the following two subsections.

\subsection{Solutions of field equations for corresponding ST theory}
This higher curvature like $F(R)$ model (in eqn.(\ref{actionF(R)})) can be transformed into scalar-tensor 
theory by using the technique discussed in \cite{tp1}. Performing a conformal transformation of the metric as 
\begin{equation}
 G_{MN}(x,\phi) \rightarrow \tilde{G}_{MN} = \exp{(\frac{1}{\sqrt{3}}\kappa\Phi(x,\phi))}G_{MN}(x,\phi),
 \label{conformal}
\end{equation}
the above action (in eqn.(\ref{actionF(R)})) can be expressed as a scalar-tensor theory with the action given by \cite{tp1} :
\begin{eqnarray}
 S&=&\int d^4x d\phi \sqrt{\tilde{G}} [\frac{\tilde{R}}{2\kappa^2} + \frac{1}{2}\tilde{G}^{MN}\partial_M\Phi \partial_N\Phi - V(\Phi)\nonumber\\
 &+&\Lambda + \exp{(-\frac{5}{2\sqrt{3}}\kappa\Phi)} V_h\delta(\phi)\nonumber\\ 
 &+&\exp{(-\frac{5}{2\sqrt{3}}\kappa\Phi)} V_v\delta(\phi-\pi),
 \label{action1ST}
\end{eqnarray}
where the quantities in tilde are reserved for the ST theory. $\tilde{R}$ is the Ricci curvature formed 
by the transformed metric $\tilde{G}_{MN}$. $\Phi(x,\phi)$ is the scalar 
field which corresponds to higher curvature degrees of freedom and $V(\Phi)$ is the scalar potential which 
for this specific choice of $F(R)$ has the form \cite{tp1},
\begin{eqnarray}
 V(\Phi)&=&\frac{1}{8\kappa^2\alpha} \exp{(-\frac{5}{2\sqrt{3}}\kappa\Phi)} 
 [\exp{(\frac{3}{2\sqrt{3}}\kappa\Phi)} - 1]^2\nonumber\\
 &-&\Lambda [\exp{(-\frac{5}{2\sqrt{3}}\kappa\Phi)} -1].
 \label{scalar_potential}
\end{eqnarray}

One can check that the above potential (in eqn.(\ref{scalar_potential})) is stable for the 
parametric regime $\alpha > 0$. The stable value ($<\Phi>$) and the mass squared ($m_{\Phi}^2$) 
of the scalar field ($\Phi$) are given by the following two equations
\begin{equation}
 \exp{(\frac{3}{2\sqrt{3}}\kappa<\Phi>)} = [\sqrt{9 - 40\kappa^2\alpha\Lambda} - 2]
 \label{vev_phi}
\end{equation}
and
\begin{equation}
 m_{\Phi}^2 = \frac{1}{8\alpha} [\sqrt{9 - 40\kappa^2\alpha\Lambda}] [\sqrt{9 - 40\kappa^2\alpha\Lambda} - 2]^{-\frac{2}{3}}.
 \label{mass_phi}
\end{equation}
Furthermore, the minimum value of the potential i.e. $V(<\Phi>)$ is non zero and serves as a cosmological constant. Thus 
the effective cosmological constant in scalar-tensor theory is $\Lambda_{eff} = \Lambda - V(<\Phi>)$ where $V(<\Phi>)$ 
is,
\begin{eqnarray}
 V(<\Phi>)&=&\Lambda + [\sqrt{9 - 40\kappa^2\alpha\Lambda} - 2]^{-\frac{5}{3}}\nonumber\\
 &[&-\Lambda + (1/8\kappa^2\alpha)[\sqrt{9 - 40\kappa^2\alpha\Lambda}-3]^2].
 \nonumber\\
\end{eqnarray}
This form of $V(<\Phi>)$ with $\Lambda < 0$ clearly indicates that $\Lambda_{eff}$ is also 
negative or more explicitly, the corresponding scalar-tensor theory for the original $F(R)$ 
model has an AdS like spacetime \cite{csaki}. Considering $\xi$ as the fluctuation 
of the scalar field over its vacuum expectation value (vev), the final form of action for 
the scalar-tensor theory in the bulk can be written as,
\begin{eqnarray}
 S&=&\int d^4x d\phi \sqrt{\tilde{G}} [\frac{\tilde{R}}{2\kappa^2} + \frac{1}{2}\tilde{G}^{MN}\partial_M\xi 
 \partial_N\xi\nonumber\\
 &-&(1/2)m_{\Phi}^2\xi^2 + \Lambda_{eff}]
 \label{action2ST}
\end{eqnarray}
where the terms up to quadratic order in $\xi$ are retained for $\kappa\xi < 1$.\\
For the case of ST theory presented in eqn.(\ref{action2ST}), $\xi$ can act as a bulk 
scalar field with the mass given by eqn.(\ref{mass_phi}). Considering a negligible backreaction 
of the scalar field ($\xi$) on the background spacetime, the solution of metric $\tilde{G}_{MN}$ 
is exactly same as RS model, i.e.,
\begin{equation}
 d\tilde{s}^2 = e^{- 2 kr_c|\phi|} \eta_{\mu\nu} dx^{\mu} dx^{\nu} - r_c^2d\phi^2 ,
 \label{grav.sol1.ST}
\end{equation}
where $k = \sqrt{\frac{-\Lambda_{eff}}{24M^3}}$ and $r_c$ is the compactification radius 
of the extra dimension in ST theory. With this metric, the scalar field equation of motion 
in the bulk is the following,
\begin{eqnarray}
 &-&\frac{1}{r_c^2}\partial_\phi[\exp{(-4kr_c|\phi|)}\partial_\phi\xi]\nonumber\\
 &+&m_{\Phi}^2\exp{(-4kr_c|\phi|)}\xi(\phi) = 0
 \label{eom.scalar.field}
\end{eqnarray}
where the scalar field $\xi$ is taken as function of extra dimensional coordinate only \cite{GW}. Considering 
non zero value of $\xi$ on branes, the above equation (\ref{eom.scalar.field}) has the general 
solution,
\begin{equation}
 \xi(\phi) = e^{2kr_c|\phi|} \big[Ae^{\nu kr_c|\phi|} + Be^{-\nu kr_c|\phi|}\big]
 \label{sol.scalar.field}
\end{equation}
with $\nu = \sqrt{4 + m_{\Phi}^2/k^2}$. Moreover $A$ and $B$ are obtained from the 
boundary conditions, $\xi(0)=v_h$ and $\xi(\pi)=v_v$ as follows :
\begin{equation}
 A = v_v e^{-(2+\nu)kr_c\pi} - v_h e^{-2\nu kr_c\pi}
 \nonumber\\
\end{equation}
and
\begin{equation}
 B = v_h (1 + e^{-2\nu kr_c\pi}) - v_v e^{-(2+\nu)kr_c\pi}.
 \nonumber\\
\end{equation}
Upon substitution the form of $A$ and $B$ into eqn.(\ref{sol.scalar.field}), one finds 
that 
\begin{eqnarray}
\xi(0)&=&{v}_h\nonumber\\
\xi(\pi)&=&v_v[1-e^{-2\nu kr_c\pi}+\frac{v_h}{v_v}e^{-(3\nu-2)kr_c\pi}]
\nonumber
\end{eqnarray}
Above values of $\xi(0)$ and $\xi(\pi)$ matches with the boundary condition (i.e. $\xi(0)=v_h$ and $\xi(\pi)=v_v$) 
by neglecting the subleading powers of $e^{-kr_c\pi}$, as have been done earlier 
by the authors in \cite{GW}.\\

\subsection{Solutions of field equations for original F(R) theory}
\begin{figure}[!h]
\begin{center}
 \centering
 \includegraphics[width=3.2in,height=2.2in]{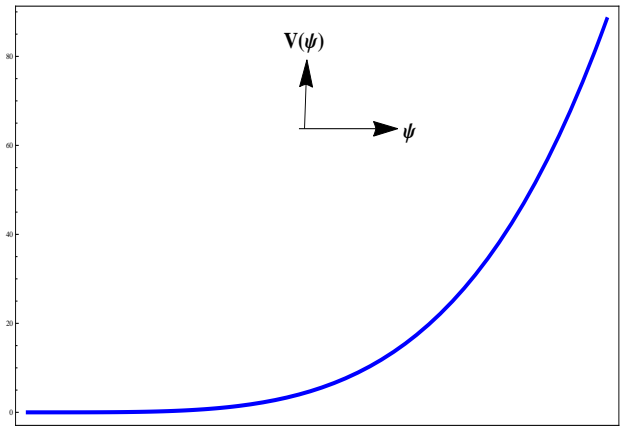}
 \caption{$V(\Psi)$ vs $\Psi$ for $M = k = 1$, $\epsilon=0.04$, $\kappa v_v = 0.01$, $\frac{v_h}{v_v} = 1.5$ and $\alpha = 10^{-22}$ (GeV)$^{-2}$.}
 \label{plot potential}
\end{center}
\end{figure}

Recall that the original higher curvature $F(R)$ model is presented by the action given in eqn.(\ref{actionF(R)}). 
Solutions of metric ($G_{MN}$) for this $F(R)$ model can be extracted from the solutions of corresponding scalar-tensor 
theory (eqn.(\ref{grav.sol1.ST}) and eqn.(\ref{sol.scalar.field})) with the help of eqn.(\ref{conformal}). Thus 
the line element in $F(R)$ model turns out to be
\begin{equation}
 ds^2 = e^{-\frac{\kappa}{\sqrt{3}}\Phi(\phi)} [e^{- 2 kr_c|\phi|} \eta_{\mu\nu} dx^{\mu} dx^{\nu} - r_c^2d\phi^2]
 \label{grav.sol1.F(R)}
\end{equation}
where $\Phi(\phi) = <\Phi> + \xi(\phi)$ and $\xi(\phi)$ is given by eqn.(\ref{sol.scalar.field}). This solution 
of $G_{MN}$ immediately leads to the separation between hidden ($\phi=0$) and visible ($\phi=\pi$) branes along 
the path of constant $x^{\mu}$ as follows :
\begin{equation}
 \pi d = r_c \int_{0}^{\pi} d\phi e^{-\frac{\kappa}{2\sqrt{3}}\Phi(\phi)} 
 \nonumber\\
\end{equation}
where $d$ is the inter-brane separation in $F(R)$ model. A fluctuation of branes around the configuration $d$ is now considered. 
This fluctuation 
can be taken as a field ($T(x)$) and this new field is assumed to be the 
function of brane coordinates only \cite{GW_radion}. Then the metric takes the following form,
\begin{eqnarray}
 ds^2 = e^{-\frac{\kappa}{\sqrt{3}}\Phi(x,\phi)} &[&e^{- 2 kT(x)|\phi|} 
 g_{\mu\nu}(x) dx^{\mu} dx^{\nu}\nonumber\\ 
 &-&T(x)^2d\phi^2]
 \label{grav.sol2.F(R)}
\end{eqnarray}
where $g_{\mu\nu}(x)$ is the induced on-brane metric and $T(x)$ is known as radion (or modulus) field. Moreover 
$\Phi(x,\phi)$ is obtained from eqn.(\ref{sol.scalar.field}) by replacing $r_c$ by $T(x)$.\\
Plugging back the solutions presented in eqn. (\ref{grav.sol2.F(R)}) into original five dimensional 
F(R) action (in eqn. (\ref{actionF(R)})) and integrating over $\phi$ yields 
the four dimensional effective action as follows \cite{tp1}
\begin{eqnarray}
 S_{eff} = \int d^4x \sqrt{-g}&\bigg[&M_{(4)}^2R_{(4)} + \frac{1}{2}g^{\mu\nu}\partial_{\mu}\Psi\partial_{\nu}\Psi\nonumber\\ 
 &-&V(\Psi)\bigg]
 \label{effective action}
\end{eqnarray}
where $M_{(4)}^2= \frac{M^3}{k}\bigg[\sqrt{9 - 40\kappa^2\alpha\Lambda} - 2\bigg]^{-5/6}$ is the 
four dimensional Planck scale, $R_{(4)}$ is the Ricci scalar formed by $g_{\mu\nu}(x)$. Moreover,  
$\Psi(x) = \sqrt{\frac{24M^3}{k}} \big[1 + \frac{20}{\sqrt{3}}\alpha k^2\kappa v_h\big] e^{-k\pi T(x)} = fe^{-k\pi T(x)}$ 
(with $f = \sqrt{\frac{24M^3}{k}} [1 + \frac{20}{\sqrt{3}}\alpha k^2\kappa v_h]$), is the 
canonical radion field and $V(\Psi)$ is the radion potential with the following form \cite{tp1}
\begin{eqnarray}
 V(\Psi)&=&\frac{20}{\sqrt{3}}\frac{\alpha k^5}{M^6} \Psi^4 
 [v_v - (v_h - \frac{\kappa v_h^2}{2\sqrt{3}}\nonumber\\ 
 &+&\frac{\kappa v_hv_v}{2\sqrt{3}})(\Psi/f)^\epsilon ]^2
 \label{potential.radion.F(R)}
\end{eqnarray}
where the terms proportional to  $\epsilon$ ($= \frac{m_{\Phi}^2}{k^2}$) 
are neglected \cite{GW,GW_radion}. 
It may be observed that $V(\Psi)$ goes to zero as $\alpha$ tends 
to zero. This is expected because for $\alpha \rightarrow 0$, 
the action contains only the Einstein part which does not produce any 
potential term for the radion field \cite{GW_radion}. 
Thus for five dimensional warped geometric model, the radion potential is 
generated from the higher order curvature term $\alpha R^2$. In figure (\ref{plot potential}), we plot 
$V(\Psi)$ against $\Psi$.
 
The potential in eqn.(\ref{potential.radion.F(R)}) has a vev at
\begin{equation}
 <\Psi> = f (\frac{v_v}{v_h})^{1/\epsilon} [1 - \frac{\kappa v_v}{2\sqrt{3}}(\frac{v_h}{v_v} - 1)]^{-1/\epsilon}
 \label{vev.radion.F(R)}
\end{equation}
as long as $\alpha>0$. Correspondingly the squared mass of radion field is as follows,
\begin{eqnarray}
 m_{rad}^2(F(R))&=&\frac{20}{\sqrt{3}}\frac{\alpha k^4}{M^6} \epsilon^2 e^{-2kd\pi} v_h^2v_v^2\nonumber\\
 &[&1 + \frac{40}{\sqrt{3}}\alpha k^2\kappa v_h] [\frac{v_h}{v_v} - 1]^2.
 \label{radion.mass.F(R)}
\end{eqnarray}
Due to the presence of $V(\Psi)$, radion field has a certain dynamics governed by 
effective field equations. In the next few sections, we examine whether the dynamics of radion field 
can trigger an inflationary scenario for the four dimensional universe or not.

\section{Solutions of effective Friedmann equations}
Considering the on-brane metric ansatz as flat FRW one i.e. 
\begin{eqnarray}
 ds_{(4)}^2&=&g_{\mu\nu}(x)dx^{\mu}dx^{\nu}\nonumber\\
 &=&dt^2 - a^2(t)\big[dx^2 + dy^2 + dz^2\big]
 \nonumber
\end{eqnarray}
where $a(t)$ is the scale factor of the visible universe. The effective field equations (obtained from the effective action 
presented in eqn.(\ref{effective action})) take the following form,
\begin{equation}
 H^2 = \frac{1}{3}\big[V(\Psi) + \frac{1}{2}(\dot{\Psi})^2\big]
 \label{eff eqn 1}
\end{equation}
and
\begin{equation}
 \ddot{\Psi} + 3H\dot{\Psi} + V'(\Psi) = 0
 \label{eff eqn 2}
\end{equation}
where an overdot denotes the derivative $\frac{d}{dt}$, $H=\frac{\dot{a}}{a}$ is known as 
Hubble parameter and the form of $V(\Psi)$ is given in eqn.(\ref{potential.radion.F(R)}). To derive 
the above equations, we assume that the radion field ($\Psi(t)$) is homogeneous in space.\\
In order to solve the effective Friedmann equations, the potential energy of radion field 
is taken as very much greater than the kinetic energy (known as slow roll approximation) i.e. 
\begin{equation}
 V(\Psi)\gg \frac{1}{2}(\dot{\Psi})^2.
 \nonumber
\end{equation}
With this approximation, eqn.(\ref{eff eqn 1}) and eqn.(\ref{eff eqn 2}) are simplified to,
\begin{equation}
 H^2 = \frac{1}{3}V(\Psi)
 \label{eff eqn 3}
\end{equation}
and 
\begin{equation}
 3H\dot{\Psi} + V'(\Psi) = 0
 \label{eff eqn 4}
\end{equation}
respectively. Substituting $H(t)$ from eqn.(\ref{eff eqn 3}) to eqn.(\ref{eff eqn 4}) and 
using the explicit form of $V(\Psi)$, we get the equation of motion for radion field 
as,
\begin{eqnarray}
 \frac{d\Psi}{dt} = -8v_v\sqrt{\frac{5}{3\sqrt{3}}\frac{\alpha k^5}{M^6}} \Psi \big(B\Psi^{\epsilon}-1\big)
 \label{eom of radion}
\end{eqnarray}
where $B=\frac{1}{v_vf^{\epsilon}}\big(v_h - \frac{\kappa v_h^2}{2\sqrt{3}} + \frac{\kappa v_vv_h}{2\sqrt{3}}\big)$. 
Eqn.(\ref{eom of radion}) immediately 
leads to the dynamics of radion field as,
\begin{eqnarray}
 &\Psi(t)=&\nonumber\\ 
 &\frac{\Psi_0}{\bigg[B\Psi_0^{\epsilon}-(B\Psi_0^{\epsilon}-1)\exp
 {\big(-8\epsilon v_v\sqrt{\frac{5}{3\sqrt{3}}\frac{\alpha k^5}{M^6}}(t-t_0)\big)}\bigg]^{1/\epsilon}},
 \label{sol of radion}
\end{eqnarray}
where $\Psi_0$ is the value of radion field ($\Psi(t)$) at $t=t_0$. Eqn.(\ref{sol of radion}) clearly indicates that 
$\Psi(t)$ decreases with time. Comparison of eqn.(\ref{vev.radion.F(R)}) and eqn.(\ref{sol of radion}) clearly reveals that the radion 
field reaches at its vev asymptotically (within the slow roll approximation) at large time ($t\gg t_0$) i.e. 
\begin{eqnarray}
 \Psi(t\gg t_0)&=&f (\frac{v_v}{v_h})^{1/\epsilon} [1 - \frac{\kappa v_v}{2\sqrt{3}}(\frac{v_h}{v_v} - 1)]^{-1/\epsilon}\nonumber\\
 &=&<\Psi>
\end{eqnarray}
This vev of radion field leads to the stabilized interbrane separation (between Planck and TeV branes) as,
\begin{eqnarray}
k\pi <T(x)> = \frac{4k^2}{m_{\Phi}^2}[\ln{(\frac{v_h}{v_v})} - \frac{\kappa v_v}{2\sqrt{3}}(\frac{v_h}{v_v} - 1)]
 \label{brane separation}
\end{eqnarray}
where $m_{\Phi}^2$ is given in eqn.(\ref{mass_phi}).\\ 
Putting the solution of $\Psi(t)$ into eqn.(\ref{eff eqn 3}) one gets, on integration, the 
evolution of scale factor as,
\begin{eqnarray}
 a(t) = C \exp{\bigg[2v_v\sqrt{\frac{5}{3\sqrt{3}}\frac{\alpha k^5}{M^6}} \big(g_1(t)-g_2(t)\big)\bigg]} ,
 \label{sol of scale}
\end{eqnarray}
where $C$ is an integration constant and $g_1(t)$ has the following form,
\begin{eqnarray}
 &g_1&(t)= - \frac{B\Psi_0^{\epsilon}}{(B\Psi_0^{\epsilon}-1)} \frac{1}{16\epsilon v_v\sqrt{\frac{5}{3\sqrt{3}}\frac{\alpha k^5}{M^6}}}\Psi_0^2*\nonumber\\ 
 &2F1&\bigg(1,1,2+\frac{2}{\epsilon},\frac{B\Psi_0^{\epsilon}}{B\Psi_0^{\epsilon}-1}
 \exp{\big(8\epsilon v_v\sqrt{\frac{5}{3\sqrt{3}}\frac{\alpha k^5}{M^6}}(t-t_0)\big)}\bigg)\nonumber\\
 &\exp&{\bigg(8\epsilon v_v\sqrt{\frac{5}{3\sqrt{3}}\frac{\alpha k^5}{M^6}}(t-t_0)\bigg)} \bigg(B\Psi_0^{\epsilon}-(B\Psi_0^{\epsilon}-1)\nonumber\\
 &\exp&{\big(-8\epsilon v_v\sqrt{\frac{5}{3\sqrt{3}}\frac{\alpha k^5}{M^6}}(t-t_0)\big)}\bigg)^{-2/\epsilon}
 \label{g1}
\end{eqnarray}

where $2F1$ symbolizes the hypergeometric function. Similarly the form of $g_2(t)$ is given by,

\begin{eqnarray}
 &g_2&(t)= - \frac{\Psi_0^{\epsilon}}{(B\Psi_0^{\epsilon}-1)} \frac{1}{16\epsilon v_v\sqrt{\frac{5}{3\sqrt{3}}\frac{\alpha k^5}{M^6}}}\Psi_0^2*\nonumber\\ 
 &2F1&\bigg(1,1,1+\frac{2}{\epsilon},\frac{B\Psi_0^{\epsilon}}{B\Psi_0^{\epsilon}-1}
 \exp{\big(8\epsilon v_v\sqrt{\frac{5}{3\sqrt{3}}\frac{\alpha k^5}{M^6}}(t-t_0)\big)}\bigg)\nonumber\\
 &\exp&{\bigg(8\epsilon v_v\sqrt{\frac{5}{3\sqrt{3}}\frac{\alpha k^5}{M^6}}(t-t_0)\bigg)} \bigg(B\Psi_0^{\epsilon}-(B\Psi_0^{\epsilon}-1)\nonumber\\
 &\exp&{\big(-8\epsilon v_v\sqrt{\frac{5}{3\sqrt{3}}\frac{\alpha k^5}{M^6}}(t-t_0)\big)}\bigg)^{1-2/\epsilon}.
 \label{g2}
\end{eqnarray}

It may be noticed from eqn.(\ref{sol of radion}) and eqn.(\ref{sol of scale}) that for $\alpha\rightarrow 0$, 
the solution of radion field and Hubble parameter become $\Psi(t)=\Psi_0$ and $H(t)=0$ respectively. It is expected 
because in the absence of higher curvature term, $V(\Psi)$ goes to zero and thus the radion field 
has no dynamics which in turn vanishes the evolution of scale factor of the universe.

\section{Beginning of inflation}
After obtaining the solution of $a(t)$ (in eqn.(\ref{sol of scale})),   
we can now examine whether this form of scale factor corresponds to an accelerating era of the early universe (i.e. $t\gtrsim t_0$)
or not. In order to check this, we expand $a(t)$ in the form of Taylor series (about $t=t_0$) and 
retain the terms only up to first order in $t-t_0$:
\begin{eqnarray}
 &a&(t\gtrsim t_0)=a_0\nonumber\\ 
 &\exp&{\bigg[2(B\Psi_0^{\epsilon}-1) \Psi_0^2 v_v\sqrt{\frac{5}{3\sqrt{3}}\frac{\alpha k^5}{M^6}}(t-t_0)\bigg]}
 \label{limiting scale}
\end{eqnarray}
where $a_0$ is the value of the scale factor at $t=t_0$ and related to the integration constant $C$ as,
\begin{equation}
 a_0 = C \exp{[-\frac{\Psi_0^2}{8}]}.
 \nonumber
\end{equation}
It is evident from eqn.(\ref{limiting scale}) that $a(t)$ corresponds to an exponential expansion at early age of the universe 
where $t_0$ specifies the onset of inflation. Moreover the Hubble parameter ($H=\frac{\dot{a}}{a}$) depends on 
the higher curvature parameter $\alpha$ and for $\alpha\rightarrow 0$, $a(t)=a_0$. 
Thus the accelerating period of the early universe is triggered entirely due to the presence of 
higher curvature term in the five dimensional bulk spacetime.

\section{End of inflation}
In the previous section, we show that the very early universe expands with an acceleration and this accelerating stage is  
termed as the inflationary epoch. In this section, we check whether the acceleration of the scale factor 
has an end in a finite time or not.\\
In the case of inflation, $\ddot{a}>0$. By relating the definition of inflation to the Hubble parameter, 
one readily obtains,
\begin{eqnarray}
 \frac{\ddot{a}}{a} = \dot{H} + H^2 > 0        
 .\label{def of inflation}
\end{eqnarray}
We now estimate the time interval which is consistent with this condition. Recall the slow roll equation 
(eqn.(\ref{eff eqn 3})) as,
\begin{eqnarray}
 H^2&=&\frac{1}{3}V(\Psi)\nonumber\\
 &=&\frac{20}{3\sqrt{3}}\frac{\alpha k^5}{M^6}v_v^2 \Psi^4 \big(B\Psi^{\epsilon}-1\big)^2.
 \nonumber
\end{eqnarray}
Differentiating both sides of this equation with respect to t, we get the time derivative 
of the Hubble parameter as follows,
\begin{eqnarray}
 \dot{H} = -\frac{160}{3\sqrt{3}} \frac{\alpha k^5}{M^6}v_v^2 \Psi^2 \big(B\Psi^{\epsilon}-1\big)^2
 \label{time der of hubble}
\end{eqnarray}
where we use the expression of $\dot{\Psi}$ from eqn.(\ref{eom of radion}). Plugging back the expressions 
of $H^2$ and $\dot{H}$ into eqn.(\ref{def of inflation}) one gets the following condition on radion field,
\begin{equation}
 \Psi > 2\sqrt{2} = \Psi_f = \Psi(t_f)
 \label{end of inflation}
\end{equation}
where $t_f$ is the time when the radion field acquires the value $2\sqrt{2}$ (in Planckian unit). 
Eqn. (\ref{end of inflation}) clearly indicates that the inflationary era of the universe continues as long as the radion field 
remains greater than $\Psi_f$ ($= 2\sqrt{2}$). Correspondingly the duration of inflation (i.e. $t_f-t_0$) can 
be calculated from the solution of $\Psi_(t)$ as follows,
\begin{eqnarray}
 \Psi_f^{\epsilon}=\frac{\Psi_0^{\epsilon}}{\bigg[B\Psi_0^{\epsilon}-(B\Psi_0^{\epsilon}-1)\exp
 {\big(-8\epsilon v_v\sqrt{\frac{5}{3\sqrt{3}}\frac{\alpha k^5}{M^6}}(t_f-t_0)\big)}\bigg]}
 \nonumber
\end{eqnarray}
Simplifying the above expression, we obtain
\begin{eqnarray}
 t_f-t_0 = \frac{1}{8\epsilon v_v\sqrt{\frac{5}{3\sqrt{3}}\frac{\alpha k^5}{M^6}}}
 \ln\bigg[\frac{\Psi_f^{\epsilon}(B\Psi_0^{\epsilon}-1)}{\Psi_0^{\epsilon}(B\Psi_f^{\epsilon}-1)}\bigg].
 \label{duration}
\end{eqnarray}
So the inflation comes to an end in a finite time. In order to estimate the duration of inflation explicitly, one needs the 
initial value of the radion field (i.e. $\Psi_0$) which can be determined from the expression of number of e-foldings, 
discussed in the next section.

\section{Number of e-foldings and Slow roll parameters}
Total number of e-foldings ($N_0$) of the inflationary era is defined as,
\begin{eqnarray}
 N_0 = \int_{t_0}^{t_f} H(t) dt.
 \label{def of e-folding 1}
\end{eqnarray}
Using the slow roll equation, the above expression is simplified to the form,
\begin{eqnarray}
 N_0 = \int_{\Psi_0}^{\Psi_f} \sqrt{\frac{V(\Psi)}{3}} \frac{1}{\dot{\Psi}} d\Psi
 \label{def of e-folding 2}
\end{eqnarray}
Putting the explicit form of $V(\Psi)$ (eqn.(\ref{potential.radion.F(R)})) 
and the time derivative of $\Psi(t)$ 
(eqn.(\ref{eom of radion})) into the right hand side of eqn.(\ref{def of e-folding 2}) and integrating over $\Psi$, 
one obtains the final result of number of e-foldings, given by
\begin{eqnarray}
 N_0 = \frac{1}{16} \big(\Psi_0^2 - \Psi_f^2\big).
 \label{e-folding}
\end{eqnarray}
It may be mentioned that the total number of e-foldings is independent of mass of the radion field (or inflaton field).\\

We may define,
\begin{eqnarray}
 N(\Psi) = N_0 - \int_{\Psi_0}^{\Psi} \sqrt{\frac{V(\Psi)}{3}} \frac{1}{\dot{\Psi}} d\Psi,
 \nonumber
\end{eqnarray}
the number of e-foldings remaining until the end of inflation when the inflaton field 
crosses the value $\Psi(t)$. Simplifying the above expression, one obtains : 
\begin{eqnarray}
 N(\Psi) = \frac{1}{16} \big(\Psi^2 - \Psi_f^2\big).
\nonumber
\end{eqnarray}

In order to test the broad inflationary paradigm as well as
particular models against precision observations \cite{Planck_result}, 
it is crucial to calculate the slow roll parameters ($\epsilon_V$ and $\eta_V$), which are defined as follows :
\begin{eqnarray}
\epsilon_V = \frac{1}{2}\bigg(\frac{V'(\Psi)}{V(\Psi)}\bigg)^2
\nonumber
\end{eqnarray}
and\\
\begin{eqnarray}
\eta_V = \bigg(\frac{V''(\Psi)}{V(\Psi)}\bigg)
\nonumber
\end{eqnarray}
The slow roll condition demands that the parameters $\epsilon_V$ and $\eta_V$ should be less than 
unity as long as the inflationary era continues. By using the form of inflaton potential ($V(\Psi)$, in eqn.(\ref{potential.radion.F(R)})), 
the above expressions can be simplified and turn out to be,
\begin{eqnarray}
\epsilon_V = \frac{8}{16N(\Psi) + \Psi_{f}^2}
\label{slow1}
\end{eqnarray}
and\\
\begin{eqnarray}
\eta_V = \frac{6}{16N(\Psi) + \Psi_{f}^2}
\label{slow2}
\end{eqnarray}

Using these expressions of slow roll parameters, 
one determines the spectral index of curvature perturbation 
($n_s$) and tensor to scalar ratio ($r$) in terms of $N_{*}$ $\big(= \frac{1}{16} \big(\Psi_{*}^2 - \Psi_f^2\big),$ 
the number of e-foldings remaining until the end of inflation when the cosmological scales exit the 
horizon and $\Psi_{*}$ is the corresponding value of the inflaton field$\big)$ \cite{habib,langlois,riotto} :
\begin{eqnarray}
 n_s = \frac{16N_* - 28}{16N_* + 8}
 \label{spectral_index}
\end{eqnarray}
and\\
\begin{eqnarray}
 r = \frac{128}{16N_* + 8}
 \label{tensor_scalar}
\end{eqnarray}
To derive eqn.(\ref{spectral_index}) and eqn.(\ref{tensor_scalar}), we use the value of $\Psi_f$ ($=2\sqrt{2}$) that 
has been obtained earlier (see eqn.(\ref{end of inflation})). From observational results ($Planck$ $2015$) \cite{Planck_result} 
$n_s$ and $r$ are constrained to be $n_s = 0.968 \pm 0.006$ and $r < 0.12$ respectively. Using eqn. (\ref{spectral_index}) and 
eqn. (\ref{tensor_scalar}), it can be easily shown that in order to make agreement between the theoretical and observational results, 
$N_*$ should be equal to $60$. Putting this value 
of $N_*$ into eqn. (\ref{spectral_index}) and eqn. (\ref{tensor_scalar}), we obtain the following results of $n_s$ and $r$ :
\begin{eqnarray}
 n_s = 0.963\nonumber\\
 r = 0.11
 \nonumber
\end{eqnarray}

Moreover, at the pivot scale ($N(\Psi)=N_*$), $\epsilon_V$ and $\eta_V$ acquire the values as $0.009$ and $0.007$ respectively.\\
In table (\ref{Table-1}), we now summarize our results.

\begin{table}[!h]
 \centering
\resizebox{\columnwidth}{1.0 cm}{%
  \begin{tabular}{|c| c| c|}
   \hline \hline
   Parameters & Theoretical results  & Observational results\\
   ($n_s$ and $r$) & from the present model (for $N_*=60$) & from $Planck$ $2015$\\ 
   \hline
   $n_s$ & 0.963 & 0.968 $\pm$ 0.006\\
   $r$ & 0.11 & $<$ 0.12\\
   \hline
  \end{tabular}%
  }
  \caption{Theoretical and observational results of $n_s$ and $r$}
  \label{Table-1}
 \end{table}
 
\begin{figure}[!h]
\begin{center}
 \centering
 \includegraphics[width=3.2in,height=2.2in]{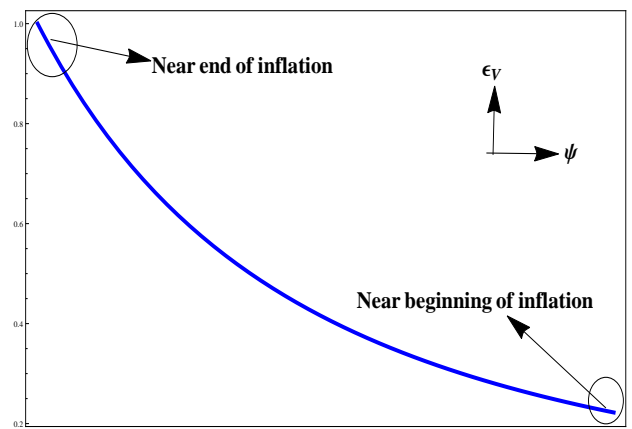}
 \caption{$\epsilon_V$ vs $\Psi$}
 \label{plot slow1}
\end{center}
\end{figure}

\begin{figure}[!h]
\begin{center}
 \centering
 \includegraphics[width=3.2in,height=2.2in]{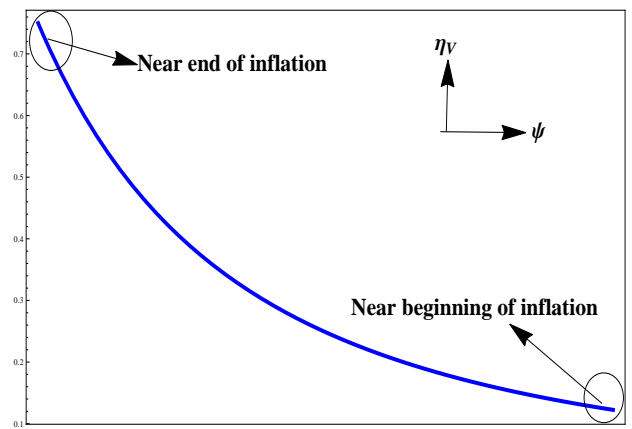}
 \caption{$\eta_V$ vs $\Psi$}
 \label{plot slow2}
\end{center}
\end{figure}

 It is evident from table (\ref{Table-1}), that the present model of five dimensional higher curvature gravity predicts the correct values for 
 $n_s$ and $r$ as per the observations of $Planck$ $2015$.\\
 However the required value of $N_*$ ($=60$) can be achieved 
if $\Psi_*$ is adjusted to the value as $\Psi_*\simeq 31$ (in Planckian unit). Also demanding the total number 
of e-foldings of inflationary era to be equal to $70$ (i.e. $N_0 = 70$), we obtain the initial value of inflaton field, 
$\Psi_0 = 33.5$ (in Planckian unit). With this value 
of $\Psi_0$, duration of inflation ($t_f-t_0$) comes as $\sim 10^{-32}$ sec (or $10^{-8}$ (GeV)$^{-1}$, see eqn.(\ref{duration})) 
if the higher curvature parameter $\alpha$ and $\kappa v_v$ are taken as
\begin{eqnarray}
 \alpha \sim 10^{-22} (GeV)^{-2}
 \label{value of alpha}
\end{eqnarray}
and
\begin{eqnarray}
 \kappa v_v = 0.01
 \nonumber
\end{eqnarray}
respectively. Furthermore, the effective gravitational constant $(M_{(4)})$ is $\sim 10^{19}$ GeV for the estimated 
value of $\alpha$ presented in eqn.(\ref{value of alpha}).\\
Once we find the initial ($\Psi_0$) and final ($\Psi_f$) values of the inflaton field, we now give the plots (figure(\ref{plot slow1}) and 
figure(\ref{plot slow2})) between the slow roll parameters and $\Psi$ (by using eqn.(\ref{slow1}) and eqn.(\ref{slow2})).

Figure (\ref{plot slow1}) and figure (\ref{plot slow2}) clearly demonstrate that as long as inflation continues, both the 
slow roll parameters ($\epsilon_V$ and $\eta_V$) remain less than unity. This behaviour of $\epsilon_V$ and $\eta_V$ 
are expected from the slow roll approximation. Furthermore, the value of $\epsilon_V$ and $\eta_V$ increase with the evolution of 
universe during the inflationary epoch and at the end of inflation, $\epsilon_V$ becomes one i.e. $\epsilon_V(\Psi_f)=1$.\\

\section{Comparison of solutions with and without slow roll approximation}
\begin{figure}[!h]
\begin{center}
 \centering
 \includegraphics[width=3.2in,height=2.2in]{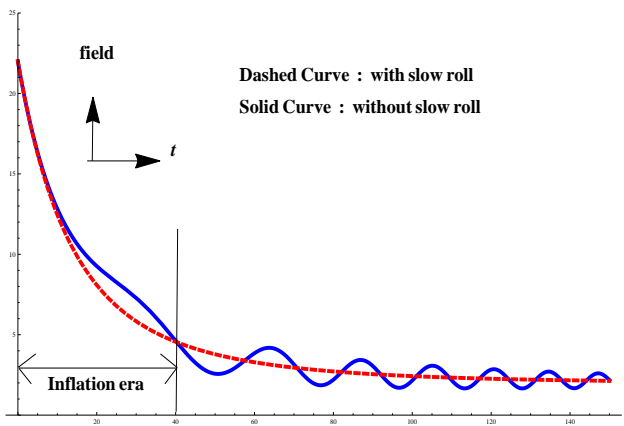}
 \caption{$\Psi(t)$ vs. $t$ with/without slow roll approximation}
 \label{plot field}
\end{center}
\end{figure}

In this section, we solve the radion field and Hubble parameter numerically from the complete form of effective 
Friedmann equations (eqn. (\ref{eff eqn 1}) and eqn. (\ref{eff eqn 2}), without slow roll approximations). These numerical 
solutions are then compared with the solutions (in eqn. (\ref{sol of radion}) and eqn. (\ref{sol of scale})) obtained 
by solving the slow roll equations.\\
Eqn. (\ref{eff eqn 1}) and eqn. (\ref{eff eqn 2}) lead to the equation of $\Psi(t)$ as follows:
\begin{eqnarray}
 \ddot{\Psi} + \sqrt{3\big[V(\Psi) + \frac{1}{2}(\dot{\Psi})^2\big]}\dot{\Psi} + V'(\Psi) = 0
 \nonumber
\end{eqnarray}
Using the form of $V(\Psi)$, above differential equation is solved numerically for $\Psi(t)$. The comparison between this numerical solution 
and the solution obtained in eqn. (\ref{sol of radion}) is presented in figure (\ref{plot field}).\\

\begin{figure}[!h]
\begin{center}
 \centering
 \includegraphics[width=3.2in,height=2.2in]{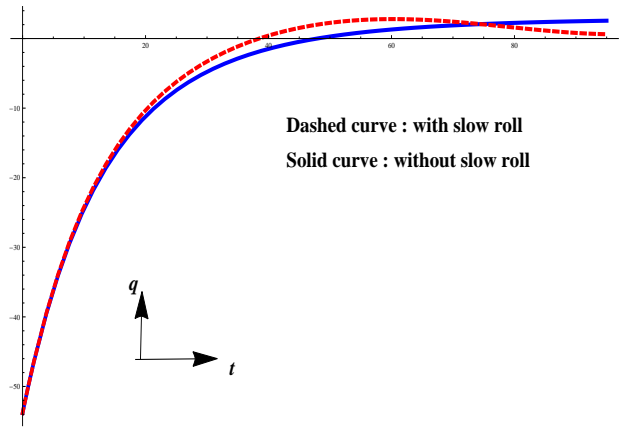}
 \caption{$q(t)$ vs. $t$ with/without slow roll approximation}
 \label{plot deceleration}
\end{center}
\end{figure}

Figure (\ref{plot field}) demonstrates that the plotted result of $\Psi(t)$ based on solving the slow roll equations 
and the plotted result of $\Psi(t)$ based on solving the full Freidmann equations (in presence of $\dot{\Psi}^2$ and $\ddot{\Psi}$) 
are almost same during the inflation. But after the inflation the acceleration term of inflaton (the term containing $\ddot{\Psi}$) 
starts to contribute and as a result the two solutions (with and without slow roll conditions) differ from each other. Moreover 
in the slow roll approximation, $\Psi(t)$ does not exhibit oscillatory phase at the end of inflation, but it tends to its minimum 
value asymptotically. Such an oscillatory character of $\Psi(t)$ occurs when the term $\ddot{\Psi}$ is taken into account 
in the equation of motion.\\
The numerical solution of Hubble parameter and correspondingly the deceleration parameter ($q(t) = -(\dot{H}+H^2)$) are also obtained 
from eqn. (\ref{eff eqn 1}). The variation of $q(t)$ versus $t$ (with/without slow roll approximation) 
is shown in figure (\ref{plot deceleration}).\\

Figure (\ref{plot deceleration}) clearly depicts that the duration of inflation predicted from the 
numerical solution of complete Friedmann equations is longer than that predicted from the solutions of slow roll equations.\\

\section{Summary and Concluding Remarks}
In this work, we consider a five dimensional compactified warped AdS model with two 
3-branes embedded within the spacetime. Due to large curvature ($\sim$ Planck scale) in the bulk, the 
spacetime is considered to be governed by higher curvature like $F(R)=R+\alpha R^2$. Our visible universe 
is identified with the TeV scale brane, which emerges out of the four dimensional effective theory. On projecting 
the bulk gravity on the brane, the extra degrees of freedom of $R_{(5)}$ appears as a scalar field on the brane and is 
known as radion field. The potential term ($V(\Psi)$) of radion field is proportional to the higher curvature parameter $\alpha$ 
and goes to zero as $\alpha\rightarrow 0$. Thus it is clear that the radion potential is generated entirely due to the presence 
of higher curvature term in the five dimensional bulk spacetime. The form of $V(\Psi)$ (in eqn.(\ref{potential.radion.F(R)})) 
indicates that the radion potential is stable as long as $\alpha$ is considered to be positive and the minimum of $V(\Psi)$ is zero.\\

From the perspective of four dimensional effective theory, we examine the possibility of ``inflationary scenario'' 
by taking the on-brane metric ansatz as a spatially flat FRW one. In the presence of radion potential,$V(\Psi)$, 
we determine the solutions (in eqn.(\ref{sol of radion}) and eqn.(\ref{sol of scale})) 
of effective Friedmann equations by considering the potential energy of radion field as very much 
greater than the kinetic energy (also known as slow roll approximation). The solution of scale factor 
corresponds to an accelerating expansion of the early universe and the rate of expansion 
depends on the parameter $\alpha$. It may be mentioned that the radion field 
as well as the scale factor become constant as $\alpha$ goes to zero. Thus it can be argued that 
due to the presence of higher curvature term, the radion field has a certain dynamics which in 
turn triggers an exponential expansion of the universe at an early epoch. The expression of duration of inflation 
($t_f-t_0$) is also obtained in eqn. (\ref{duration}) which reveals that the accelerating phase of the universe 
terminates within a finite time.\\

We determine the slow roll parameters ($\epsilon_V$ and $\eta_V$) and it is found that both $\epsilon_V$ and $\eta_V$ remain less than 
unity as long as the inflation continues. The expressions of slow roll parameters yield the spectral index of curvature perturbation 
($n_s$) and tensor to scalar ratio ($r$) in terms of $N_*$ (number of e-foldings remaining until the end of inflation from the pivot scale). 
For $N_*=60$, $n_s$ and $r$ take the values as 
$n_s=0.963$ and $r=0.11$, which match with the observational results based on the observations of $Planck$ $2015$ (see table (\ref{Table-1})). 
Thus our considered model of five dimensional higher curvature gravity predicts the correct values of $n_s$ and $r$ as per the observations 
of $Planck$ $2015$. Moreover the duration of inflation comes as $\sim 10^{-32}$ sec (or $10^{-8}$ (GeV)$^{-1}$) 
if the higher curvature parameter $\alpha$ is of the order of $10^{-22}$ (GeV)$^{-2}$.\\

Finally we find the solution for the radion field and Hubble parameter numerically 
from the complete form of Friedmann equations (without the slow roll approximations). During the inflation, these numerical solutions 
are almost same with the solutions of slow roll equations, as demonstrated in figure (\ref{plot field}) and figure (\ref{plot deceleration}). 
Another important point to note is that in the slow roll approximation, $\Psi(t)$ does not exhibit oscillatory phase 
at the end of inflation, but it tends to its minimum value asymptotically, while such an oscillatory behaviour of inflaton 
is indeed there if the slow roll approximation is relaxed, i.e., the 
acceleration term of $\Psi(t)$ in the equation of motion is not dropped, as depicted in figure (\ref{plot field}).

\end{document}